\documentclass[12pt,preprint]{aastex}

\usepackage{amssymb}
\usepackage{graphicx}
\usepackage{subfigure}

\shorttitle{The Effect of WISE/GALEX for SED fitting}
\shortauthors{Z. Fan and S. Wang}

\begin{document}

\title{The Effects of the WISE/{\it GALEX} Photometry for the SED-Fitting with M31 Star Clusters and Candidates}

\author{Zhou Fan\altaffilmark{1} \& Song Wang\altaffilmark{1}}

\altaffiltext{1}{Key Laboratory for Optical Astronomy, National
  Astronomical Observatories, Chinese Academy of Sciences, 20A Datun
  Road, Chaoyang District, Beijing 100012, China}
\email{zfan@bao.ac.cn}

\begin{abstract}
Spectral energy distribution (SED) fitting of stellar population 
synthesis models is an important and popular way to constrain the physical
parameters ---e.g., the ages, metallicities, masses for stellar population 
analysis. The previous works suggest that both blue-bands and red-bands  
photometry works for the SED-fitting. Either blue-domained or red-domained 
SED-fitting usually lead to the unreliable or biased results. Meanwhile, it 
seems that extending the wavelength coverage could be helpful. Since 
the Galaxy Evolution Explorer ({\it GALEX})
and Wide-field Infrared Survey Explorer (WISE) provide the FUV/NUV and 
mid-infrared $W1$/$W2$ band data, we extend the SED-fitting to a wider 
wavelength coverage. In our work, we analyzed the effect of adding the 
FUV/NUV and $W1$/$W2$ band to the optical and 
near-infrared $UBVRIJHK$ bands for the fitting with the Bruzual \& Charlot 
2003 (BC03) models and {\sc galev} models. It is found that the FUV/NUV bands 
data affect the fitting results of both ages and metallicities much more 
significantly than that of the WISE $W1$/$W2$ band with the BC03 models. 
While for the {\sc galev} models, the effect of the WISE $W1$/$W2$ band 
for the metallicity fitting seems comparable to that of  {\it GALEX} FUV/NUV 
bands, but for age the effect of the $W1$/$W2$ band seems 
less crucial than that of the FUV/NUV bands. Thus we conclude that the 
{\it GALEX} FUV/NUV bands are more crucial for the SED-fitting of ages and 
metallicities, than the other bands, and the high-quality UV data (with high 
photometry precision) are required. 
\end{abstract}


\section{Introduction}
\label{intro.sec}

SED-fitting of the simple stellar population (SSP) models is an important 
method to estimate the physical parameters, e.g., the ages, metallicities and 
masses of star clusters by the $\chi^2_{\rm min}$ techniques. It is based on 
the precise multi-band photometry and it has been applied in great number of 
recent works for the star clusters in the extra-galaxies. \citet{rdg03} fit 
the SEDs from broad-band Ultra-voilet (UV), optical to
NIR observations of $Hubble$ $Space$ $Telescope$ ($HST$) to derive age,
metallicity and extinction of star cluster system of NGC 3310.
\citet{bas05} constrain the age, mass, extinction, and effective radius of 
1152 star clusters in M51 by fitting the SEDs from the $HST$ imaging 
from ultraviolet to near infrared, with 
Galaxy Evolutionary Synthesis Models \citep[{\sc galev};][]{lf06,kot09}, 
which considers the gaseous emission lines as well as continuum emission and 
thus it is important for the young star clusters.
Based on the Beijing-Arizona-Taiwan-Connecticut (BATC) multi-color photometry 
system, \citet{fan06,ma07,ma09,ma11,ma12,wang10,wang12} have done series of 
work on the SED-fitting of M31 star clusters with the SSP models such as 
\citet{bc03} models and {\sc galev} models. In order to improve the fitting 
results and partly break the age-metallicity degeneracy, other photometry 
bands, such as the $UBVRI$ broadband, Two Micron All Sky Survey (2MASS) $JHK$ 
band, Galaxy Evolution Explorer ({\it GALEX}) near-ultraviolet (NUV) and 
far-ultraviolet (FUV), even the Sloan Digital Sky Survey (SDSS) $ugriz$ bands 
are applied for the fitting, especially for the UV passband, 
  \citep[see, e.g.,][]{kav07,bia09}. \citet{fan10,fd12,fd14} also fit the 
SEDs of M31 and M33 star clusters in the $UBVRIJHK$ bands and $ugriz$ band 
with \citet{bc03} models/{\sc galev}/BS-SSP models and 
{\sc parsec isochrones}. 
Similarly, \citet{kang12} performed the photometry with {\it GALEX} 
NUV and FUV imaging data of star clusters and fitted the SED in up to 16 
passbands ranging from FUV to NIR with data gathered from the literature 
values and 
the Revised Bologna Catalog (RBC v4) and they obtained ages and masses of 176 
young ($\le1$ Gyr) clusters and 446 old ($>1$ Gyr) clusters.
 
Since WISE provides near- and mid-infrared W1/W2 (3.4$\mu$/4.6$\mu$) band data 
respectively, we could extend to a wider wavelength coverage for the
SED-fitting and check the effect of the WISE bands. 
As a matter of fact, in the similar bands, $Spitzer$- Infrared Array 
Camera (IRAC) colors has been applied for the stellar population analysis. 
For instance, \citet{pele02} investigated the ([3.6]-[4.5]) color for the 
stellar populations in the early-type galaxies and found that the color is 
likely to be a good metallicity indicator: the color becomes bluer with 
increasing metallicity, which is attributed to the increasing importance 
of a CO absorption feature in the [4.5] bandpass. \citet{bj12} also 
analyzed the $Spitzer$ IRAC colors with several stellar population synthesis 
models for the massive, old GCs in M31 and it is found that although the 
colors become slightly bluer with age, the effect is quite small, just a few 
tenths of a magnitude. Further, the [3.6]-[4.5] color dose not show apparent
relation with metallicity either for the models or for the cluster data.
\citet{mei12} found that the independent component analysis (ICA) 
technique with the [3.6]-[4.5] color can isolate the old stellar light from 
contaminant emission in a sample of six disk galaxies. After removing the 
emission from evolved red objects with low mass-to-light ratios, it is found
that the underlying old distribution of light with [3.6]-[4.5] colors is 
consistent with the colors of K and M giants. \citet{mei14} also found that
the the [3.6]-[4.5] color is bluer at higher metallicity due to the CO 
absorption in the [4.5] band for the giant stars, but it is not very 
sensitive to age. \citet{nor14} also found that a linear relation between 
the $W1-W2$ color and metallicity for the nearby GCs and the early-type 
galaxies although the scatter is large and only the models considering the 
effect of the increasing CO absorption in the W2 band can successfully 
reproduce the observed trend. The $W1$-$W2$ color is insensitive to age 
for age $>2$ Gyr. It is also found that the mass-to-light ratio
$M/L$ for old stellar population at 3.6 $\mu m$ varies modestly with the 
age and metallicity \citep[see, e.g.,][]{que15}, which is also confirmed 
by \cite{rock15}, who found that both [3.6]-[4.5] color and $W1$-$W2$ 
become only 0.01-0.02 mag redder with ages for ages above 2 Gyr and 
it becomes up to 0.04 mag bluer with increasing metallicity. 
\citet{nor16} found that the $W1$ band is an exceptional tracer of stellar 
mass for the quiescent/early-type galaxies and it is highly recommended.

In fact, the effects and comparisons of different combinations of passbands
for the SED fitting have been studies in many previous works. \citet{and04} 
investigate effects caused by the number of passbands, different
passband combinations, observational errors and non-continuous models, by
fitting the SEDs of artificial star clusters with evolutionary 
synthesis models with different set and number of passbands from $UBVRIJH$. 
They found that the $U$ and $B$ band are most important, and $V$ and 
near-infrared are also helpful for the fitting the age, metallicity, 
extinction and mass. As we mentioned above, \citet{rdg03} also investigated 
the effects by fiting the SEDs of NGC 3310 star clusters with different 
passband combinations from UV to NIR observations of $HST$ and they found 
that the blue-selected passband combinations lead to a slightly bias towards 
lower ages but the red-dominated passband combinations, especially dominated 
by NIR filters, should be avoided. \citet{rdg05} analyzed the systematic 
uncertainty of the SED-fitting with $HST$ imaging observations in 
UV+optical+NIR band and conclude that as extensive a wavelength coverage 
as possible is required to obtain robust age and mass estimates for the 
SED-fittings of various models with reasonable uncertainties. 
\citet{kg07} also fit different passband combination of the optical and 
NIR photometry with BC03 and Maraston models and compared the stellar+gas mass 
with the dynamical mass in different class of galaxies.

In this paper, we focus on the analysis and comparison of the parameters 
of age and metallicity, which could be derived directly from the SED-fit, 
while the mass is estimated with mass-to-light ratio $M/L$ and it depends on 
the estimated age and metallicity, which introduce the uncertainties again. 
Besides, the number of works for constraining masses of M31 star clusters is
relatively less than that of ages and metallicities (even no masses included
in the RBC catalog), which makes it more difficult to compare with, although 
WISE $W1$ is an exceptional tracer of mass after all.
We compare the results of SED-fit with different combinations of the 
photometry bands, from {\it GALEX} FUV, NUV, to the JHK and WISE bands of the 
M31 star clusters. This paper is organized as follows. In
Section \ref{mod.sec} we describe the stellar population synthesis models 
applied in our work and the convolution of the AB magnitudes. In Section 
\ref{fit.sec}, we introduce our sample of M31 GCs and the fitting methods. 
In Section \ref{res.sec} we give the fitting results and comparisons of the 
ages and metallicities based on $\chi^2_{\rm min}$ fitting, using various 
models and methods. Finally, we summarize our work and give the conclusions 
in Section \ref{sum.sec}.

\section{The Models and $\chi_{min}^2$-fitting Method}
\label{mod.sec}

In our work, two stellar population synthesis models are used.

1. The \citet[][hereafter BC03]{bc03} stellar population synthesis models
provide SEDs of various physical parameters, such as ages, metallicities and 
masses. The stellar evolutionary tracks of Padova 1994 and 2000 Padova are 
given, and the Initial Mass Functions (IMFs) of \citet{sal} and \citet{chab03}
IMFs are provided. The wavelength ranges are from 91{\AA} to 160 $\mu$m. For
the Padova 1994 tracks, models of metallicities for $Z=0.0001$,
0.0004, 0.004, 0.008, 0.02, and 0.05 are provided, as for the
Padova 2000 tracks, models of metallicities ($Z=0.0004$,
0.001, 0.004, 0.008, 0.019, and 0.03) are given. Since the metallicity steps
are too large for the fitting, we interpolate the models to attain smaller 
intervals of the parameter space, i.e. 51 metallicities with equal steps in 
logarithmic space) to obtain the subtle results. Meanwhile, 221 ages 
0-20 Gyr in unequally spaced time steps are provided. Different combinations 
of Padova 1994/2000 stellar evolutionary tracks and IMFs are computed for the 
models. However, it is known that for a different IMF dose not affect the 
results including the uncertainties more significantly than the stellar 
evolutionary track dose. We should note that the best-fitting metallicity 
range for the Padova 2000 tracks is not as wide as that obtained from the
Padova 1994 tracks due to the metallicity limitations of the models.

2. The {\sc galev} models,
which can be applied to constrain the chemical evolution of the
gas and the spectral evolution of the stellar population in star clusters
or galaxies simultaneously. The stellar evolutionary tracks/isochrones 
Padova and Geneva are provided, and in our work we adopt the Padova 
evolutionary track. The models not only give the photometry but also provide 
the Lick absorption-line indices for different stellar populations with
different star-formation rates or even the SSPs for the single burst. 
For the ages, 5001 values 4 Myr - 20 Gyr provide by the models, and 
metallicities of $Z=0.0001$, 0.0004, 0.001, 0.004, 0.008, 0.02, and 0.05
are given, for which the grid step is too large. Thus we also interpolate the 
metallicities to a grid of 51 values which lead to more accurate results.
The model spectra coverages the wavelength from XUV at $\sim90$ {\AA} to 
the FIR 160 $\mu$m, with a spectral resolution of 20 {\AA} in the 
UV-optical and 50-100 {\AA}A in the NIR wavelength range. Then we convolve the 
model spectra with the filter transmissions and obtain the model magnitudes. 
Nevertheless as \citet{fd12} pointed out, the {\sc galev} models usually 
predict younger ages than other models, like BC03 and works better for 
young stellar populations. 

In fact, \citet{nor14} found that many models, including the BC03 models
and the {\sc galev} models, fail to fit the observed $W1$-$W2$ colors of 
stellar populations dramatically around solar metallicity. The observed 
scatter is too large to make the $W1$-$W2$ color to be one metallicity 
indicator. These two models can give the correct zeropoint in the $W1$ band, 
however they substantially underpredict the absolute zeropoint in 
the $W2$ band. \cite{rock15} also suggest that the {\sc galev} models 
predict redder color by about 0.10-0.13 of absolute values of the Spitzer 
([3.6]-[4.5]) than that of their SSP models due to the theoretical stellar 
atmospheres and not considering the CO absorption in the 4.5 $\mu m$. 
Fortunately it is also well known that in continuing the trend for IR 
photometry, the WISE $W1$ and $W2$ bands have significantly reduced 
sensitivity to age and metallicity (see the discussion in 
Sect.~\ref{intro.sec}). Therefore it seems that the determination of age 
and metallicity is unaffected by the use of WISE photometry for the BC03 and
{\sc galev} models.

For both of the {\sc galev} and BC03 SSP models, the theoretical spectra can 
be convolved to magnitudes in the AB system using the filter-response functions
in FUV/NUV/$UBVRIJHK$ bands and $W1$/$W2$ bands \citep{jar11}. The AB 
magnitudes of synthesis models are given by,
\begin{equation}
  m_{\rm AB}(t)=-2.5~{\log~\frac{\int_{\lambda_1}^{\lambda_2}{{\rm d}\lambda}~{\lambda}~F_{\lambda}(\lambda,t)~R(\lambda)}{\int_{\lambda_{1}}^{\lambda_{2}}{{\rm d} \lambda}~{\lambda}~R(\lambda)}}-48.60,
  \label{eq1}
\end{equation}
where $R(\lambda)$ is filter-response function and
$F_{\lambda}(\lambda,t)$ is the flux, which is a function of
wavelength ($\lambda$) and evolutionary time ($t$). $\lambda_1$ and
$\lambda_2$ are the lower and upper wavelength cutoffs of the
respective filter.

\section{The Cluster Sample Selection and the $\chi_{min}^2$-Fitting}
\label{fit.sec}

In our work, we collected the photometry of M31 star clusters and candidates 
from ultraviolet bands to the middle infrared bands for the SED-fitting. For 
the photometry of FUV and NUV bands, i.e., Galaxy Evolution Explorer 
({\it GALEX}) data, the $UBVRI$ broad band data and Two Micron All Sky 
Survey (2MASS) $JHK$ band data are from Revised Bologna Catalogue of M31 GCs 
and candidates \footnote{http://www.bo.astro.it/M31/} \citep[RBC v5,][]
{gall04,gall06,gall09}. Since the catalog also includes the non-cluster 
objects, such as stars or background galaxies, which may contaminate our 
fitting results, we then exclude these objects from the catalog and only 
include the star clusters and candidates in our sample, namely $f=$1, 2 or 8 
in RBC. In the catalog, the photometry only from the ultraviolet bands to 
the near-infrared bands (i.e., FUV and NUV bands of {\it GALEX} data, the 
$UBVRI$ band and 2MASS $JHK$ bands) are included. The RBC catalog dose not 
include the middle infrared photometry. Fortunately, All WISE Source Catalog 
of Infrared Processing and Analysis Center (IPAC) Infrared Science Archive 
(IRSA)\footnote{http://irsa.ipac.caltech.edu/Missions/wise.html} provides
the WISE \citep{wri} profile-fit photometry and curve-of-growth corrected 
``standard-aperture'' photometry in {\it W1}, {\it W2}, {\it W3}, {\it W4} 
bands, of which the central wavelengths are 3.4, 4.6, 12 and 22 $\mu$m. 
Since the profile-fit photometry provides the most accurate measurements for 
unresolved objects, we adopted it for our SED-fitting.
The standard deviation of WISE magnitudes 
(limiting magnitudes) of 11 frames for the S/N of 5 are 17.11, 15.66, 11.40, 
and 7.97 mag in {\it W1}, {\it W2}, {\it W3}, {\it W4} bands \citep{wri}. 
It suggests that the sensitivity of {\it W3}, {\it W4} bands are not high 
enough for the fitting of our M31 star cluster sample. Therefore, 
Table~\ref{t1.tab} only lists the WISE photometry associated with the 
uncertainties in {\it W1}, {\it W2} bands, which are actually applied in 
our work. For the convenience of the fitting and comparisons, we only select 
the star clusters and candidates in RBC with the available photometry in all 
bands, namely from {\it GALEX} FUV, NUV, broadband $UVBRI$, 2MASS $JHK$ as 
well as the WISE {\it W1}, {\it W2} bands. Finally we have only 123 star 
clusters and candidates in our sample. The photometry of WISE {\it W1} and
{\it W2} bands are listed in Table~\ref{t1.tab}, which are in the the Vega 
system.

The {\it GALEX} FUV, NUV data of the RBC are actually from the literature 
works of \citet{rey07} and \citet{kang12}. For the former work, the authors 
applied the DAOPHOT II package \citep{stet87} to perform the photometry in 
both FUV and NUV bands within a radius of 3 pixel ($4''.5$) for each point 
source. In FUV and NUV bands, 5-16 and 19-44 isolated stars per frame 
were applied for the aperture corrections. \citet{kang12} adopted the same 
photometry method and parameters as done by \citet{rey07}. While for the 
WISE {\it W1}, {\it W2}, the azimuthally averaged PSF with FWHMs of $6''.1$ 
and $6''.4$ \citep{wri}. Thus, although the background of the galaxy is 
complicated and pixel scale is relatively large, the background estimated 
is just in a very small region, slightly larger than the photometry 
aperture. Therefore, it is relatively uniform for the background flux 
subtraction and it dose not affect much of the photometry accuracy, which 
also can be seen from the photometry errors.

For the convenience of model SED-fitting, we convert all the photometry of 
Vega system to AB system in our sample for the bands from $UBVRIJHK$ and 
{\it W1}, {\it W2} bands using the \citet{kuru} SEDs.
As we know, reddening values could affect SED fitting significantly. 
Although \citet{sf11} recalibrate the infrared-based dust map of 
\citet{sfd98} with SDSS photometry to higher accuracy even 
outside of the SDSS footprint, they only consider the Galactic extinction.
For reddening correction of M31 star clusters, and the reddening values 
of M31 galaxy should be considered as well.
In our work, reddening values for our sample star clusters were adopted from
\citet{cw11} and \citet{fan08}, in higher priority of former work 
as their reddenings of star clusters were derived from spectroscopy. For 
the unavailable ones not found in the literature works above, we adopted 
$E(B-V)=0.24$ mag instead, which is the average reddening value of 
\citet{cw11}, as the representative reddening value of M31 
star clusters. The extinction $A_{\lambda}$ can be computed using
the equations of \citet{ccm89}, and we adopted a typical foreground
Milky Way extinction law, $R_V = 3.1$. We fitted the SEDs using
\begin{equation}
  \chi^2_{\rm min}={\rm
    min}\left[\sum_{i=1}^{8}\left({\frac{M_{\lambda_i}^{\rm
	  obs}-M_{\lambda_i}^{\rm mod}(t,\rm [Z/H]}
      {\sigma_{M,i}}}\right)^2\right],
  \label{eq2}
\end{equation}
where $M_{\lambda_j}^{\rm mod}(t,\rm [Z/H])$ is the
$i^{\rm th}$ magnitude provided in the stellar population model for
age $t$, metallicity $\rm [Z/H]$; $M_{\lambda_i}^{\rm obs}$ represents the 
observed dereddened magnitude in the $i^{\rm th}$ band.

Eq.~\ref{eq3} represents the errors associated with our SED-fittings,
\begin{equation}
  \sigma_{M,i}^{2}=\sigma_{{\rm obs},M,i}^{2}+\sigma_{{\rm mod},M,i}^{2},
  \label{eq3}
\end{equation}
where $\sigma_{M,i}$ is the magnitude uncertainty in the $i^{\rm th}$
filter. For the photometric errors of RBC, \citet{gall04} suggested 
for the typical error of CCD photometry are 0.08 mag in $U$, 0.05 mag in 
$BVRI$, 0.1 mag in $J$, and 0.2 mag in $HK$; for the photographic magnitudes
the photometric error is 0.05-0.2 mag. Actually in the updated version of 
RBC v5, a series of high precision photometry have been included. In our 
sample, most of the photometry of RBC are from \citet{bh00,bh01} and 
\citet{fan10} and we adopted the photometric errors from these 
literature works
if available. For those photometric errors which can not be found  
in the literature works, we adopted a mean photometric errors depending 
on the brightness of the sources. The model errors adopted were 0.05
mag, which is the typical photometric error for the \citet{bc03} and 
{\sc galev} SSP models \citep[e.g.,][]{fan06,ma07,ma09,wang10,fd14}.

\section{The Fitting Results and Discussion}
\label{res.sec}

In order to check the fitting results of our work when adding the WISE data, 
we compare that with the results from various literature works. 
Figure~\ref{fig1}
shows the comparisons of metallicity fitted with Padova 2000 evolutionary 
track and \citet{chab03} IMF of \citet{bc03} models and that from four 
of recent works. \citet{cbq} have determined the 
metallicities, ages and masses of 306 star clusters in M31, which are selected
from the Large Sky Area Multi-Object Fibre Spectroscopic Telescope (LAMOST) 
spectral survey \citep{zhao12}. The metallicities derived from full 
spectral fitting of PEGASE-HR 
model and \citet{vaz} model, as well as that derived from Lick Fe indices and 
$\rm EZ\_Ages$ code are given in \citet{cbq}. However we adopted the 
metallicities of PEGASE-HR model for the comparison, which seems most 
reliable and most comprehensive, and it is shown on the top left panel.
The dashed line represents the best linear fit and the slope is 0.45, 
indicating a positive correlation between our SED-fitting result of BC03 model
agree with that of \citet{cbq}, with the systematic offset of 
$\overline{\rm [Fe/H]_{Chen+2016} - [Fe/H]_{our~work}}=0.06\pm0.65$ dex.
The Top Right Panel shows the comparison between our fitting result and 
that of \citet{cw11}, who have taken the high-quality 
spectra of 323 old M31 star cluster with the 6.5-m MMT telescope and they have
determined the $\rm [Fe/H]$ by using the MW GC bi-linear relation and the 
Lick Fe indices. It seems the correlation between the metallicities 
from \citet{cw11} and our result is weak, with the linear fit  
slope (dashed line) of only 0.14, and systematic offset is 
$\overline{\rm [Fe/H]_{Caldwell+2011} - [Fe/H]_{our~work}}=0.01\pm0.53$ dex.
Besides, \citet{kang12} provide a compiled catalog by collecting the 
metallicity measurements of 399 star clusters in M31 from the spectroscopic 
observations of \citet{cw11}, \citet{gall09}, \citet{per02} and \citet{bh00}. 
The mean value of metallicity from the literature values is adopted in 
the catalog. However, for star clusters with a metallicity value in only one 
literature work, the value is adopted. The comparison is shown on the 
bottom left panel, suggesting that the metallicities of \citet{kang12} is 
slightly lower than that of our results, with systematic offset of
$\overline{\rm [Fe/H]_{Kang+2012} - [Fe/H]_{our~work}}=-0.04\pm0.68$ dex.
The slope of the linear fit is $-0.02$, suggesting that there is almost no 
correlation between results of the two works, which may be due to the 
systematic offsets between literature values gathered in their final 
catalogue.
In addition, \citet{fan10} updated the $UBVRI$ photometry and in order to 
determine the ages and masses of 445 confirmed globular-like and candidate 
clusters of M31, the spectroscopic metallicities are collected from 
\citet{per02}, \citet{bh00} and \citet{hbk91}, which are shown on the bottom 
right panel. It suggests that the metallicities from the compiled catalogue 
of \citet{fan10} is systematically higher than the value of our work, with the 
systematic offset of $\overline{\rm [Fe/H]_{Fan+2010} - [Fe/H]_{our~work}}=0.35\pm0.75$ dex. The slope of the linear fit is 0.67, indicating that their
result is consistent with that of our work basically, at least showing a
positive trend for the correlation.
  
Further the {\sc galev} models of \citet{kro} IMF is also applied for the 
same comparisons as Figure~\ref{fig1}, shown in Figure~\ref{fig2}. It seems
that the agreement of our results and that from literature works are 
better overall, which can be seen from the dashed lines (the slopes of linear 
fits). The systematic offsets are $0.17\pm0.51$, $0.11\pm0.39$, 
$0.07\pm0.54$ and $0.46\pm0.60$ and the slopes are 0.89, 0.67, 0.49 and 1.08 
respectively in the order of literature works as Figure~\ref{fig1}. We 
find that the systematic offset of \citet{kang12} is the smallest while that 
of \citet{fan10} is the largest, but seems all can be ignored if considering 
the errors; for the slope, it is found that \citet{kang12}
is the worst fit while \citet{fan10} is the best. In fact the it can be 
seen that our fit results basically agree with the metallicities of 
\citet{cbq} and \citet{cw11}.

Figure~\ref{fig3} is the same as Figure~\ref{fig1} but for comparisons of the 
ages, derived from SED-fitting with Padova 2000 evolutionary track and 
\citet{chab03} IMF of \citet{bc03} model and that from literature works: 
\citet{cbq} obtained the ages through full-spectral fitting the LAMOST data 
with PEGASE-HR models / Vazdekis models. Although the $\rm EZ\_Ages$ code 
and SED-fitting of $ugriz$ bands also have been applied for the age estimates, 
the ages from PEGASE-HR models seems most reliable and are adopted for the 
comparison as discussed above, which is shown on the Top Left Panel. It is 
found for the most old star clusters, i.e., age ${\rm log}~t>\sim9.5$ (yr), 
the agreement is roughly good, while for the clusters younger than that,
our fits gives younger ages. The dashed line is the best linear fit and 
the slope is 0.01. It can be seen that the ages of \citet{cbq} seems 
in a consistent range between ${\rm log}~t\sim9.5$ and $\sim10.3$ (yr). 
As we discussed above, \citet{cw11} determined the ages of a sample of M31 GCs 
with the $\rm EZ\_Ages$ and the Lick indices from MMT spectra. However, for 
clusters which fall outside the index - index grids, the ages are set to 14 
Gyr. The comparison is shown on the Top Right Panel, from which it is found 
that most ages from the fitting of \citet{cw11} are at the upper limit. 
The slope of the linear fit is $-0.04$ and it can be seen that as most of the 
ages are at the upper limit, which makes it seem that the ages from 
\citet{cw11} are almost constant, independent of our result.
\citet{kang12} estimate the ages of 182 
young clusters (younger than 1 Gyr) by multi-band SED fitting, which is 
shown on the Bottom Left Panel. It seems that the ages of \citet{kang12} are
systematically younger ($\sim0.3$dex) than that of our estimates. The 
slope of the best fit is 0.57, suggesting that the agreement is not good, 
but the trend is positive.
The ages of 445 confirmed globular-like and candidate clusters from 
\citet{fan10} are determined by $\chi^2_{min}-$fitting of the $UBVRIJHK$ 
photometry and the BC03 models, for which the comparison are shown on the 
Bottom Right Panel. The slope of the best linear fit is 0.59, and we 
found that the ages derived from \citet{fan10} are younger than our results 
for ${\rm log}~t>\sim10$ (yr).

Similar to Figure~\ref{fig3}, we also plot the age comparison of 
{\sc galev} models of \citet{kro} IMF in Figure~\ref{fig4}. The slopes of the 
best linear fits are 0.20, $-0.02$, 0.60 and 1.24 respectively in the order of
literature works as Figures above. We find that for the \citet{fan10} 
the agreement is the best, while for \citet{cw11} and \citet{cbq} the agreement
seems not good and ages from literature works seems independent of our 
fit ages, which is may 
be due to the same reason as Figure~\ref{fig3}. While for the ages of 
\citet{kang12}, our result is basically agree with the literature values.

Before compare the different sets of photometry passbands, we would like to 
know that the effects of different stellar evolutionary tracks and different 
IMFs. We compared the metallicity derived from 
different stellar evolutionary tracks and IMFs of the \citet{bc03} models in 
Figure~\ref{fig5}. The photometry of all bands are used for the fittings on 
all the panels. The Top Left Panel shows the comparisons of Padova 
1994 and Padova 2000 evolutionary tracks with same IMF of \citet{sal}, and 
the comparison of fitting results due to the two different tracks with same 
IMF of \citet{chab03} are shown on the Top Right Panel. It can be seen
that for the metallicity $\rm [Fe/H]>\sim-1.3$ dex, the metallicities derived 
from two evolutionary tracks basically consistent with each other, although
there are few outliers with $\rm [Fe/H]>0$ dex in the fitting of Padova 1994 
track where the Padova 2000 fits seems systematically lower. 
However, since the lower limit of the metallicity of the two tracks are 
different, i.e. $\rm [Fe/H]=-2.2490$ for Padova 1994 and $\rm [Fe/H]=-1.6469$ 
for Padova 2000, the difference between the results derived from the two 
evolutionary tracks becomes significant, especially for metallicity 
close to the lower limit of Padova 2000 evolutionary tracks. 
It can be seen that the upper two panels are almost in 
the same case and the IMF almost dose not affect the fits. Similarly, 
the comparisons of different IMFs \citet{sal} and \citet{chab03} 
with Padova 1994 evolutionary tracks are shown on Bottom Left Panel; the 
same comparison but with Padova 2000 evolutionary tracks are shown on Bottom 
Right Panel. Apparently the metallicities derived from IMFs of \citet{sal} 
and \citet{chab03} agree with each other very well for any case, {which 
also suggests that the IMF dose not affect models significantly.

Figure~\ref{fig6} is the same as Figure~\ref{fig5} but for the ages, which are 
also derived from evolutionary tracks of Padova 1994/Padova 2000 on the top 
panels and IMFs of \citet{chab03}/\citet{sal} of the \citet{bc03} models on 
the bottom panels. The photometry of all bands are used for the fitting. 
From the top panels we can see that the results derived from the two different 
evolutionary tracks, Padova 1994 or the Padova 2000, basically consist 
with each other for both young and old ages, except for some outlier which 
seems older in Padova 1994 models but younger in Padova 2000 models. 
The dashed lines are the best linear fit, showing the difference of results 
fit by two evolutionary tracks. On the bottom panels, again we found that the 
results from IMFs of \citet{chab03} and \citet{sal} agree with each other very 
well, although there are some outliers.
It suggests that for the \citet{bc03} models, the IMFs of \citet{chab03} and
\citet{sal} do not significantly affect the fitting results, either for the 
Padova 1994 or the Padova 2000 evolutionary tracks. Since the \citet{chab03} 
IMF and Padova 1994 evolutionary track are more up-to-date, we will apply them 
in the following work.

In order to figure out the effects of the FUV, NUV and $W1$, $W2$ in the
SED-fitting, we compare the results with different sets of photometry
passband combinations. Table~\ref{t2.tab} is the Ages and Metallicities Derived
from the SED $\chi_{min}^2$-fitting with BC03 models \citep{bc03} of Padova
2000 stellar evolutionary track and IMF of \citet{chab03}, which is more
up-to-date. The three cases are considered in our work: 1. fitting with
all bands (FUV, NUV, $UBVRIJHK$, $W1$, $W2$); 2. fitting without WISE data
(FUV, NUV, $UBVRIJHK$); 3. fitting without {\it GALEX} data ($UBVRIJHK$, $W1$,
$W2$). Figure~\ref{fig7} presents the comparisons of metallicities fitted with
\citet{bc03} models and photometry in all bands and that without 
{\it GALEX} data on the Left Panel or the fitting without WISE data on the 
Right Panel. Obviously, it found that the fitting without WISE data 
agree significantly better with the fitting of all band than the case without 
{\it GALEX} data as the scatter is much smaller. It may 
indicate that effect of the {\it GALEX} data plays an much more significantly
role than that of WISE data, or say, the SED-fitting is much more sensitive 
to the {\it GALEX} data than that of WISE data. Therefore the precision of the 
{\it GALEX} UV bands is important for the fitting results. Figure~\ref{fig8} 
is the same but shows the comparisons of ages from with photometry of all 
bands and that without {\it GALEX} data on the Left Panel and that without 
WISE data on Right Panel. Again, the ``best'' assumptions, Padova 2000 
evolutionary track and \citet{chab03} IMF are applied in the fitting.
Similarly, we found that the ages from fitting without WISE data agree much
better with fitting of all band than the case without {\it GALEX} data. 
However it is worth noting that some outliers around ${\rm log}~t\sim9.2$ 
fitted without WISE data but ${\rm log}~t\sim10.2$ fitted with all-band or 
${\rm log}~t\sim10.3$ fitted without WISE data but ${\rm log}~t\sim9.2$ fitted 
with all-band, which seems lead to ``unstable'' results. We have check the fit
carefully and found the photometry and fits are good. It may be due to that 
the models have very small distance between the two parameter node. Further 
we also can see that the number of the points is quite few compared to the 
whole sample. Actually it is found that the agreement is good in general.
Thus we conclude that the {\it GALEX} data is much more sensitive to the 
SED-fittings than the case without WISE data in the \citet{bc03} models.
In addition, we found that for the clusters with age ${\rm log}~t>\sim10$ 
(yr) for the all-band fitting, the results from fitting without {\it GALEX} 
data seems systematically younger, while for clusters $\rm \sim9<log~t<\sim9.5$
the results agree with each other well although there are some outliers likely 
to be older than that fit with all-band data.

Furthermore, we also would like to see the effect of the different IMF for the
{\sc galev} models. Figure~\ref{fig9} is the same as Figure~\ref{fig5} 
but for comparisons of metallicity fitted with {\sc galev} models.
The fitting results of \citet{kro}/Scalo/\citet{sal} IMFs are compared.
We found that the fitting results of the three IMFs are basically the same, 
suggesting that the effect on different IMF for the {\sc galev} models are
quite slight on metallicity, which is even can be ignored. Figure~\ref{fig10} 
is the same as Figure~\ref{fig6} but showing ages derived models with 
different IMFs of \citet{kro}/Scalo/\citet{sal}. The comparisons shows that 
the IMFs seems almost do not affect the fitting results either. However, we 
prefer the IMF of \citet{kro}, which is more up-to-date and reasonable. 

We also would like to check the effect of the $GALEX$ FUV and NUV bands and 
that of the WISE $W1$ and $W2$ bands for SED-fitting with the {\sc galev} 
models. Table~\ref{t3.tab} is the same as Table~\ref{t2.tab}, but for the 
ages and Metallicities derived from the SED fitting with {\sc galev} models
and Kroupa IMF. All the three cases are considered: 1. fitting with all bands
(FUV, NUV, $UBVRIJHK$, $W1$, $W2$); 2.fitting without WISE data (FUV, NUV,
$UBVRIJHK$); 3. fitting without {\it GALEX} data ($UBVRIJHK$, $W1$, $W2$).
Figure~\ref{fig11} is the same as Figure~\ref{fig7} but for the 
{\sc galev} models of \citet{kro} IMF. We compared the fitting metallicities 
with photometry in all bands and that without {\it GALEX} data on the Left 
Panel, and the comparison of that without WISE data on the Right Panel. 
The dashed lines are the best linear fits. We found that 
the results basically consist with each other for both left and right panels,
despite of scatters and a few outliers. For the left panel, it is found 
that the fitting without {\it GALEX} data will lead to higher metallicity 
around $\rm [Fe/H]\sim-1$ in the all-band fit. While the right panel shows 
that the metallicity fitted without WISE data seems higher than that fitted 
with all-band photometry for $\rm [Fe/H]<\sim-1$.

Figure~\ref{fig12} is the same as Figure~\ref{fig8} but for ages fitted with 
the {\sc galev} models of \citet{kro} IMF. It shows the fitting results from 
{\sc galev} models \citet{kro} IMF with all-band photometry and the fitting 
results without {\it GALEX} data on the Left Panel and the fitting results 
without WISE data on the Right Panel. The dashed lines are the best 
linear fit and it seems the ages derived without {\it GALEX} is systematically 
older than that from all-band fit by $\sim0.2$ dex. However the ages fitted 
without WISE data agree well with that of all-band fit on the right panel. 
Thus it is found that the fitting results are much more sensitive to the FUV 
and NUV bands of {\it GALEX} than that of the WISE $W1$ and $W2$ bands, since 
the agreement of fitting results is much better for the latter than that for 
the former. 
On the other hand, the uncertainties of ages increase significantly when 
the FUV and NUV band data are involved in the fitting, indicating that the 
high-quality FUV and NUV data is much more important for the age-fitting than 
the $W1$ and $W2$ bands. In other words, introducing the WISE $W1$ and $W2$ 
bands data to the SED-fitting is helpful but not as significant for the age 
fitting of the {\sc galev} models as that of the FUV and NUV data.

\section{Summary and Conclusion}
\label{sum.sec}

In our work, we collected the photometry of M31 star clusters and candidates 
from ultraviolet bands to the middle infrared bands for the SED-fitting.
For the photometry, the FUV and NUV bands, $UBVRI$ broadband, 2MASS $JHK$ bands
are from RBC v5 catalog. The WISE $W1$/$W2$ band photometry are downloaded 
from the IPAC/IRAS website. The $\chi_{min}^2$ technique is applied for the 
SED-fitting.

The \citet{bc03} models with Padova 2000 track and \citet{chab03} IMF are 
adopted in the fitting, which are more up-to-date. First we compare
our fitting results with the recent works of \citet{cbq}, \citet{cw11}, 
\citet{kang12} and \citet{fan10}, which give the fitting results of 
large samples of star clusters.

1. For the metallicity, our fitting result agree with that of \citet{cbq}
in general, with the systematic offset of $0.06\pm0.65$ dex. While the 
correlation between the metallicities from \citet{cw11} and our 
result seems weak, but the systematic offset is only $0.01\pm0.53$ dex, which
shows good consistency at the average level. The 
metallicities of \citet{kang12} is slightly lower than that of our results, 
with systematic offset of $-0.04\pm0.68$ dex. The metallicities of 
\citet{kang12} is almost independent of our fits, which may be due to the 
systematic offsets of metallicities that the authors gathered from various 
literature works. While for comparison of \citet{fan10}, it suggests 
that the 
metallicities are systematically higher than that of our work, with the 
systematic offset of $0.35\pm0.75$ dex and their result is consistent with 
that of our work in general. We also compared with {\sc galev} models and
it is found that the metallicities from literature works agree with 
our fitting results better than that of \citet{bc03} models overall, 
especially for \citet{cw11} and \citet{cbq}.

2. For the comparison of ages from literature works, we found the results
of both \citet{cw11} and \citet{cbq} are almost independent of our results 
derived from either \citet{bc03} models or the {\sc galev} models, which may 
be due to the reason that the ages of the two literature works are 
concentrated in 
a very narrow range or at an upper limit in the fitting. However for the 
results of \citet{kang12} and \citet{fan10}, the best linear fit shows 
positive correlations and agreements with our results in general for both
models.

We compared the effects of the different IMFs and evolutionary tracks for 
the SED-fitting:

1. For the \citet{bc03} models, it is found that fitted metallicities of the 
models with Padova 1994 and the Padova 2000 evolutionary tracks are 
basically consistent with each other for $\rm [Fe/H]>\sim-1.3$ dex with the 
IMF of either \citet{sal} or \citet{chab03}, despite of systematic offset 
for few outliers with $\rm [Fe/H]>0$. However for the 
metallicity close to the lower limits of the Padova 2000 tracks, 
the difference between the two fitting resuts becomes significant. It is also
noted that for either Padova 1994 and the Padova 2000 evolutionary track, 
different IMF dose not affect the result significantly. While for ages, the 
results derived from the evolutionary tracks of Padova 1994 and Padova 2000 
agree with each other better than that for metallicity, except for a few 
outliers. Again different IMF seems not affect the fitted ages significantly
for either Padova 1994 or Padova 2000 evolutionary tracks. 

2. For the {\sc galev} models, the different IMFs have been compared in the 
SED-fitting. We found that the effects of different IMFs for \citet{kro}/
Scalo/\citet{sal} are quite insignificant for either metallicity-fitting or
the age-fitting.

Further we investigated the fitting results of different passband combinations.
Three main cases are considered in our analysis: 1. fitting with all bands
(FUV, NUV, $UBVRIJHK$, $W1$, $W2$); 2. fitting without WISE data (i.e., FUV,
NUV, $UBVRIJHK$); 3. fitting without {\it GALEX} data (i.e., $UBVRIJHK$, $W1$,
$W2$). Two different SSP models, \citet{bc03} and {\sc galev} are applied in
the comparison and analysis:

1. For the \citet{bc03} models with Padova 2000 tracks, for either ages or
metallicities, the fitting results without WISE data agree with that of 
all-band data better than the case fitting without {\it GALEX} data and 
fitting with all-band data. Thus we conclude that the {\it GALEX} data is 
more sensitive to the SED-fittings than the WISE data with the \citet{bc03} 
models. 

2. For the {\sc galev} models with IMF of \citet{kro}, for metallicity,
the fitting results either without {\it GALEX} data or without WISE data 
agree with all-band fitting results in general. However, for the fitting of 
ages, {\it GALEX} data seems affect the fitting results more significantly 
than the WISE data.

Therefore, it is found that {\it GALEX} FUV and NUV bands play more 
important roles for the fitting than that of WISE $W1$ and $W2$ bands and 
the observing accuracy of FUV and NUV bands are more crucial than that of 
the mid-infrared bands, such as $W1$ and $W2$.

\acknowledgements

The authors thank the anonymous referee for helpful suggestions that greatly 
improved the manuscript.
This research was supported by the National Program on Key Research and 
Development Project (Grant No. 2016YFA0400804), the National Key Basic 
Research Program of China (973 Program) grant 2015CB857002, and the 
National Natural Science Foundation of China (NFSC) through grants 11373003, 
11603035 and U1631102.

\appendix		   %%appendicial material is supported

\clearpage
\pagestyle{empty}
\begin{deluxetable}{rrrrrr}
  \tablecolumns{6} \tablewidth{0pc} \tablecaption{The WISE Photometry in W1/W2
    bands, which is downloaded from All Wise Source Catalog of IPAC Infrared 
    Science Archive (IRSA), http://irsa.ipac.caltech.edu/Missions/wise.html. 
    We only list the W1/W2 bands, which are reliable for the fittings.\label{t1.tab}}
  \tablehead{
    \colhead{Name} & \colhead{W1}  & \colhead{W2} &
    \colhead{Name} & \colhead{W1}  & \colhead{W2} \\
    \colhead{} & \colhead{(mag)} & \colhead{(mag)} &
    \colhead{} & \colhead{(mag)} & \colhead{(mag)} } 
  \startdata
  B003 & $  15.05\pm   0.04$ & $  15.03\pm   0.07$ &
B009 & $  14.32\pm   0.03$ & $  14.77\pm   0.06$ \\
B010 & $  13.88\pm   0.03$ & $  13.63\pm   0.03$ &
B011 & $  13.67\pm   0.03$ & $  14.09\pm   0.04$ \\
B012 & $  12.69\pm   0.02$ & $  12.73\pm   0.02$ &
B019 & $  11.91\pm   0.02$ & $  11.95\pm   0.02$ \\
B020 & $  12.12\pm   0.02$ & $  12.18\pm   0.02$ &
B022 & $  15.19\pm   0.04$ & $  15.50\pm   0.09$ \\
B023 & $  10.64\pm   0.02$ & $  10.67\pm   0.02$ &
B025 & $  13.94\pm   0.06$ & $  13.76\pm   0.04$ \\
B027 & $  13.07\pm   0.04$ & $  13.11\pm   0.03$ &
B038 & $  13.62\pm   0.04$ & $  13.77\pm   0.04$ \\
B046 & $  15.20\pm   0.04$ & $  15.35\pm   0.08$ &
B047 & $  14.94\pm   0.03$ & $  14.97\pm   0.06$ \\
B048 & $  13.04\pm   0.04$ & $  13.34\pm   0.07$ &
B049 & $  14.88\pm   0.10$ & $  14.16\pm   0.05$ \\
\enddata
\end{deluxetable}

\clearpage
\pagestyle{empty}
\begin{deluxetable}{rr|rr|rr|rr}
%  \rotate
  \tablecolumns{8} \tablewidth{0pc} \tablecaption{Ages and Metallicities
    Derived from the SED fitting with BC03 models of Padova 1994 stellar 
    evolutionary track. The IMFs of \citet{sal} and \cite{chab03} are adopted.
    \label{t2.tab}}
  \tablehead{
    \colhead{}& \colhead{}& \multicolumn{2}{c}{All Data} & \multicolumn{2}{c}{Without WISE Data} & \multicolumn{2}{c}{Without {\it GALEX} Data}\\
    \cline{3-8} 
    \colhead{No.} & \colhead{Name} & \colhead{$\rm [Fe/H]$} &
    \colhead{$\rm Age$} & \colhead{$\rm [Fe/H]$} & 
    \colhead{$\rm Age$} & \colhead{$\rm [Fe/H]$} &
    \colhead{$\rm Age$} \\
    \colhead{} & \colhead{} & \colhead{(dex)} & \colhead{(Gyr)} &
    \colhead{(dex)} & \colhead{(Gyr)} & 
    \colhead{(dex)} & \colhead{(Gyr)}} \startdata
    1 &    B003 & $   -1.63_{ -0.00}^{+  0.11} $ & $  12.250_{-1.540}^{+ 3.230} $ 	 & $   -1.63_{ -0.00}^{+  0.10} $ & $  12.250_{-1.381}^{+ 2.593} $ 	 & $   -1.61_{ -0.02}^{+  0.29} $ & $   8.750_{-3.839}^{+ 8.600} $  \\
  2 &    B009 & $   -1.53_{ -0.10}^{+  0.13} $ & $  10.500_{-0.798}^{+ 1.119} $ 	 & $   -1.53_{ -0.10}^{+  0.12} $ & $  10.500_{-0.631}^{+ 1.118} $ 	 & $   -1.57_{ -0.06}^{+  0.29} $ & $  14.250_{-9.171}^{+ 5.750} $  \\
  3 &    B010 & $   -0.98_{ -0.06}^{+  0.05} $ & $   1.434_{-0.007}^{+ 0.030} $ 	 & $   -1.63_{ -0.00}^{+  0.05} $ & $  17.000_{-3.027}^{+ 3.000} $ 	 & $   -0.64_{ -0.18}^{+  0.12} $ & $   1.700_{-0.102}^{+ 0.495} $  \\
  4 &    B011 & $   -0.85_{ -0.05}^{+  0.05} $ & $   1.434_{-0.005}^{+ 0.017} $ 	 & $   -0.84_{ -0.05}^{+  0.04} $ & $   1.434_{-0.007}^{+ 0.018} $ 	 & $   -0.58_{ -0.15}^{+  0.14} $ & $   2.750_{-0.631}^{+ 2.506} $  \\
  5 &    B012 & $   -1.44_{ -0.08}^{+  0.08} $ & $   1.900_{-0.101}^{+ 0.129} $ 	 & $   -1.07_{ -0.04}^{+  0.05} $ & $   1.434_{-0.013}^{+ 0.073} $ 	 & $   -1.63_{ -0.00}^{+  0.28} $ & $   3.500_{-1.369}^{+ 4.929} $  \\
  6 &    B019 & $   -1.38_{ -0.05}^{+  0.04} $ & $  20.000_{-1.044}^{+ 0.000} $ 	 & $   -1.38_{ -0.04}^{+  0.04} $ & $  20.000_{-0.889}^{+ 0.000} $ 	 & $   -0.84_{ -0.13}^{+  0.14} $ & $  12.000_{-4.718}^{+ 8.000} $  \\
  7 &    B020 & $   -1.41_{ -0.05}^{+  0.04} $ & $  20.000_{-1.034}^{+ 0.000} $ 	 & $   -1.42_{ -0.04}^{+  0.04} $ & $  20.000_{-0.933}^{+ 0.000} $ 	 & $   -0.82_{ -0.15}^{+  0.15} $ & $  12.000_{-5.561}^{+ 8.000} $  \\
  8 &    B022 & $   -0.84_{ -0.12}^{+  0.10} $ & $   1.278_{-0.031}^{+ 0.006} $ 	 & $   -1.32_{ -0.11}^{+  0.13} $ & $   2.000_{-0.163}^{+ 0.245} $ 	 & $   -1.03_{ -0.20}^{+  0.22} $ & $   1.278_{-0.042}^{+ 0.012} $  \\
\enddata
\end{deluxetable}

\clearpage
\pagestyle{empty}
\begin{deluxetable}{rr|rr|rr|rr}
%  \rotate
  \tablecolumns{8} \tablewidth{0pc} \tablecaption{Ages and Metallicities
    Derived from the SED fitting with {\sc galev} models and Kroupa IMF. 
    \label{t3.tab}}
 \tablehead{
   \colhead{}& \colhead{}& \multicolumn{2}{c}{All Data} & \multicolumn{2}{c}{Without WISE Data} & \multicolumn{2}{c}{Without {\it GALEX} Data}\\
    \cline{3-8}
    \colhead{No.} & \colhead{Name} & \colhead{$\rm [Fe/H]$} &
    \colhead{$\rm Age$} & \colhead{$\rm [Fe/H]$} &
    \colhead{$\rm Age$} & \colhead{$\rm [Fe/H]$} &
    \colhead{$\rm Age$} \\
    \colhead{} & \colhead{} & \colhead{(dex)} & \colhead{(Gyr)} &
    \colhead{(dex)} & \colhead{(Gyr)} &
    \colhead{(dex)} & \colhead{(Gyr)}} \startdata   
   1 &    B003 & $   -1.70_{ -0.00}^{+  0.21} $ & $   2.692_{-0.543}^{+ 0.460} $ 	 & $   -1.70_{ -0.00}^{+  0.22} $ & $   2.692_{-0.467}^{+ 0.410} $ 	 & $   -1.70_{ -0.00}^{+  0.09} $ & $  15.912_{ -9.441}^{+  0.088} $ \\
  2 &    B009 & $   -1.52_{ -0.17}^{+  0.15} $ & $   3.268_{-0.438}^{+ 0.398} $ 	 & $   -1.34_{ -0.15}^{+  0.14} $ & $   2.976_{-0.413}^{+ 0.375} $ 	 & $   -1.70_{ -0.00}^{+  0.35} $ & $   1.220_{ -0.093}^{+ 14.071} $ \\
  3 &    B010 & $   -1.38_{ -0.15}^{+  0.12} $ & $   1.792_{-0.182}^{+ 0.142} $ 	 & $   -1.45_{ -0.16}^{+  0.14} $ & $   1.836_{-0.147}^{+ 0.125} $ 	 & $   -1.56_{ -0.14}^{+  0.16} $ & $   4.912_{ -3.085}^{+ 11.088} $ \\
  4 &    B011 & $   -0.95_{ -0.08}^{+  0.06} $ & $   1.372_{-0.143}^{+ 0.231} $ 	 & $   -0.95_{ -0.07}^{+  0.06} $ & $   1.304_{-0.077}^{+ 0.254} $ 	 & $   -1.13_{ -0.20}^{+  0.18} $ & $   3.312_{ -1.458}^{+  7.784} $ \\
  5 &    B012 & $   -1.70_{ -0.00}^{+  0.07} $ & $   1.764_{-0.118}^{+ 0.102} $ 	 & $   -1.70_{ -0.00}^{+  0.12} $ & $   1.820_{-0.101}^{+ 0.090} $ 	 & $   -1.70_{ -0.00}^{+  0.08} $ & $   1.592_{ -0.082}^{+  0.237} $ \\
  6 &    B019 & $   -0.92_{ -0.06}^{+  0.04} $ & $   1.416_{-0.167}^{+ 0.151} $ 	 & $   -0.92_{ -0.05}^{+  0.04} $ & $   1.428_{-0.173}^{+ 0.147} $ 	 & $   -1.17_{ -0.14}^{+  0.12} $ & $  15.256_{ -9.958}^{+  0.744} $ \\
  7 &    B020 & $   -0.95_{ -0.05}^{+  0.04} $ & $   1.340_{-0.096}^{+ 0.176} $ 	 & $   -0.95_{ -0.05}^{+  0.04} $ & $   1.340_{-0.093}^{+ 0.165} $ 	 & $   -1.20_{ -0.13}^{+  0.15} $ & $  15.956_{-11.315}^{+  0.044} $ \\
  8 &    B022 & $   -1.70_{ -0.00}^{+  0.07} $ & $   1.848_{-0.232}^{+ 0.580} $ 	 & $   -1.70_{ -0.00}^{+  0.11} $ & $   1.948_{-0.177}^{+ 0.522} $ 	 & $   -1.70_{ -0.00}^{+  0.08} $ & $   1.000_{ -0.266}^{+  0.052} $ \\
\enddata
\end{deluxetable}

\begin{figure}
  \resizebox{\hsize}{!}{\rotatebox{0}{\includegraphics{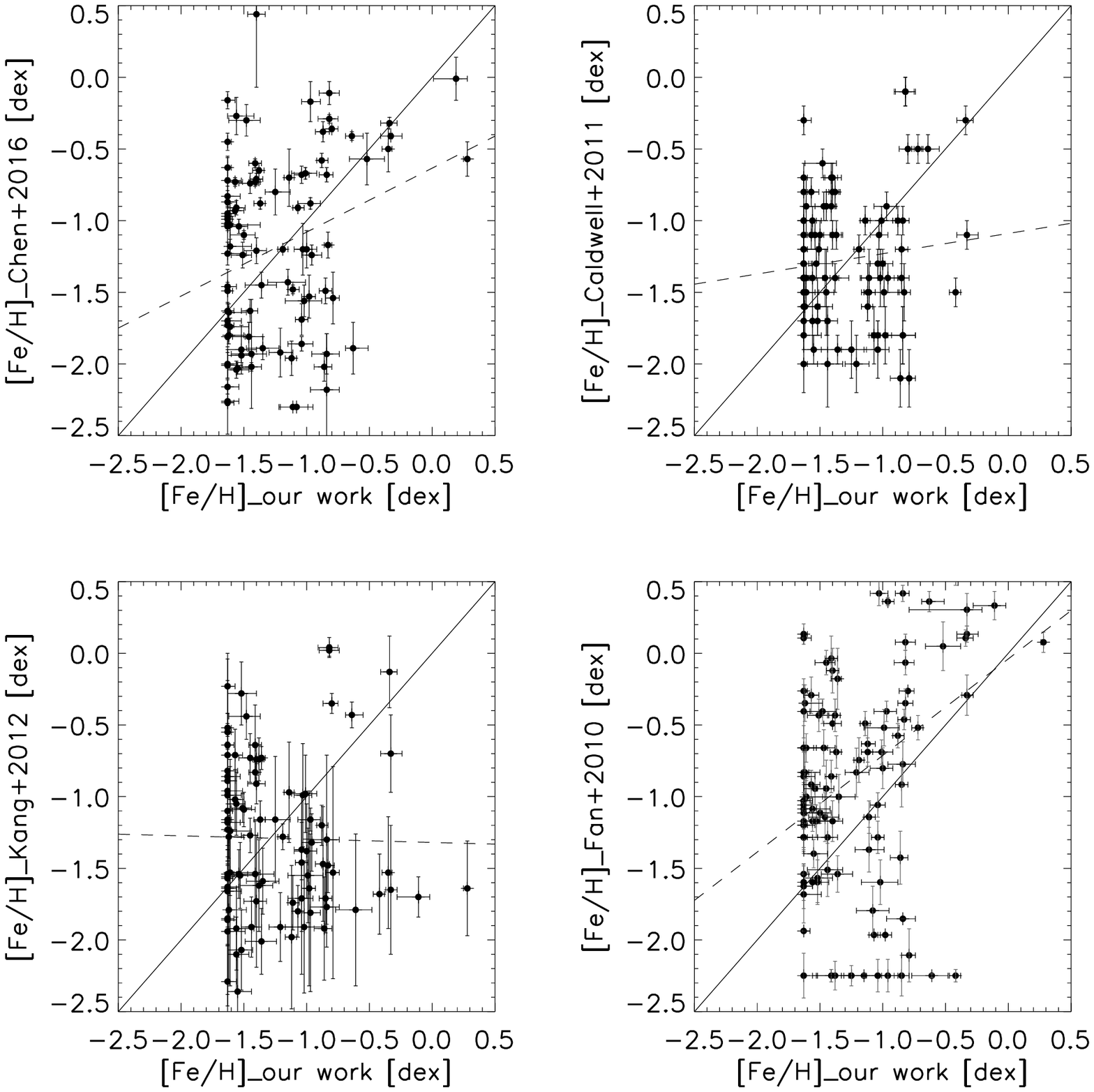}}}
  \caption{Comparisons of metallicity fit with Padova 2000 evolutionary 
    track and \citet{chab03} IMF of \citet{bc03} model and that from 
    literature works: \citet{cbq} (\sc Top Left Panel); \citet{cw11} (\sc Top 
    Right Panel); PBH spectroscopies \citep{per02,bh00,hbk91} 
    (\sc Bottom Left Panel); \citet{fan10} (\sc Bottom Right Panel).
    The dashed lines represent the best linear fits.}
  \label{fig1}
\end{figure}

\begin{figure}
  \resizebox{\hsize}{!}{\rotatebox{0}{\includegraphics{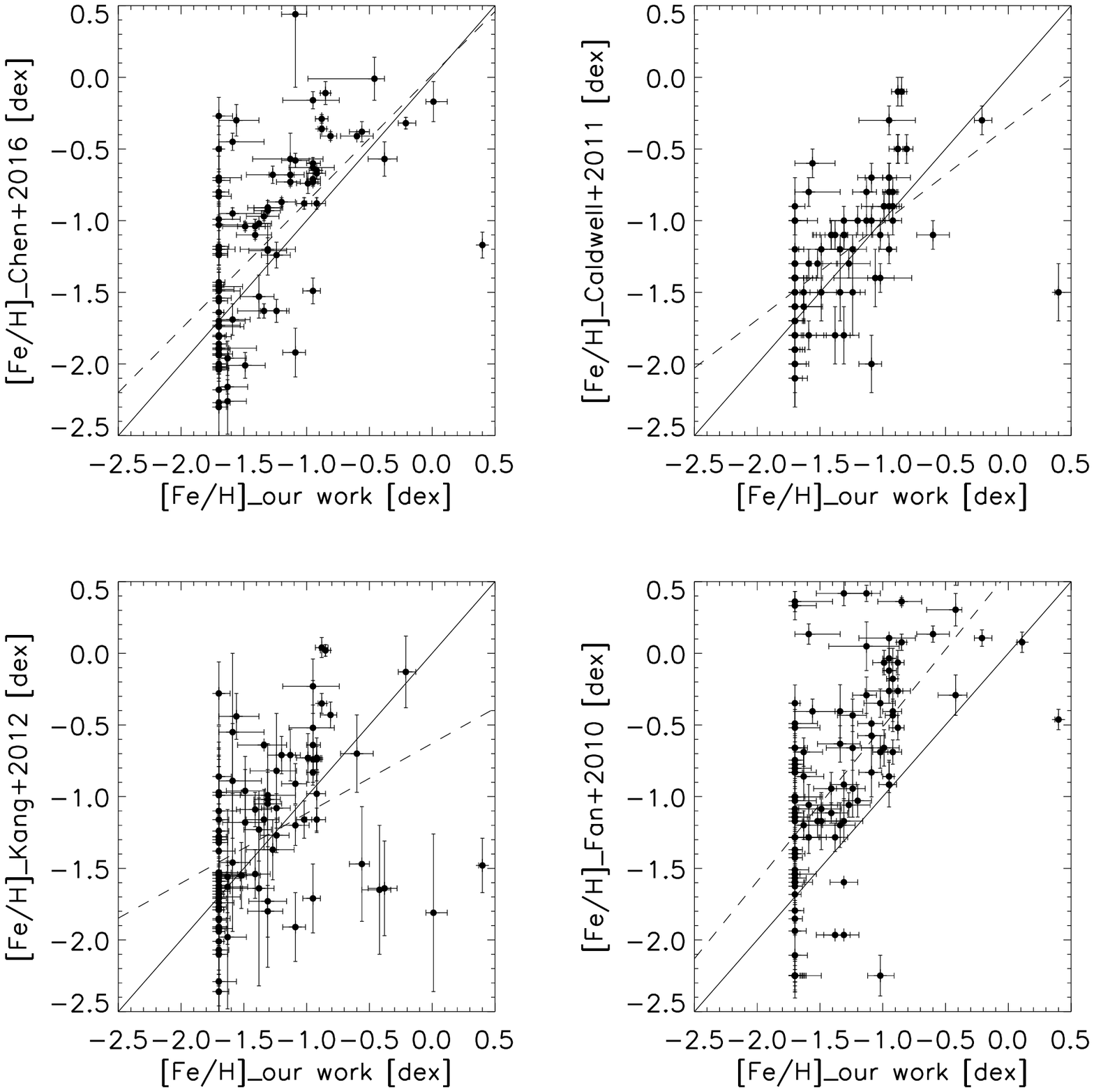}}}
  \caption{The same as Figure~\ref{fig1} but for {\sc galev} models of 
    Kroupa IMF. The dashed lines represent the best linear fits.}
  \label{fig2}
\end{figure}

\begin{figure}
  \resizebox{\hsize}{!}{\rotatebox{0}{\includegraphics{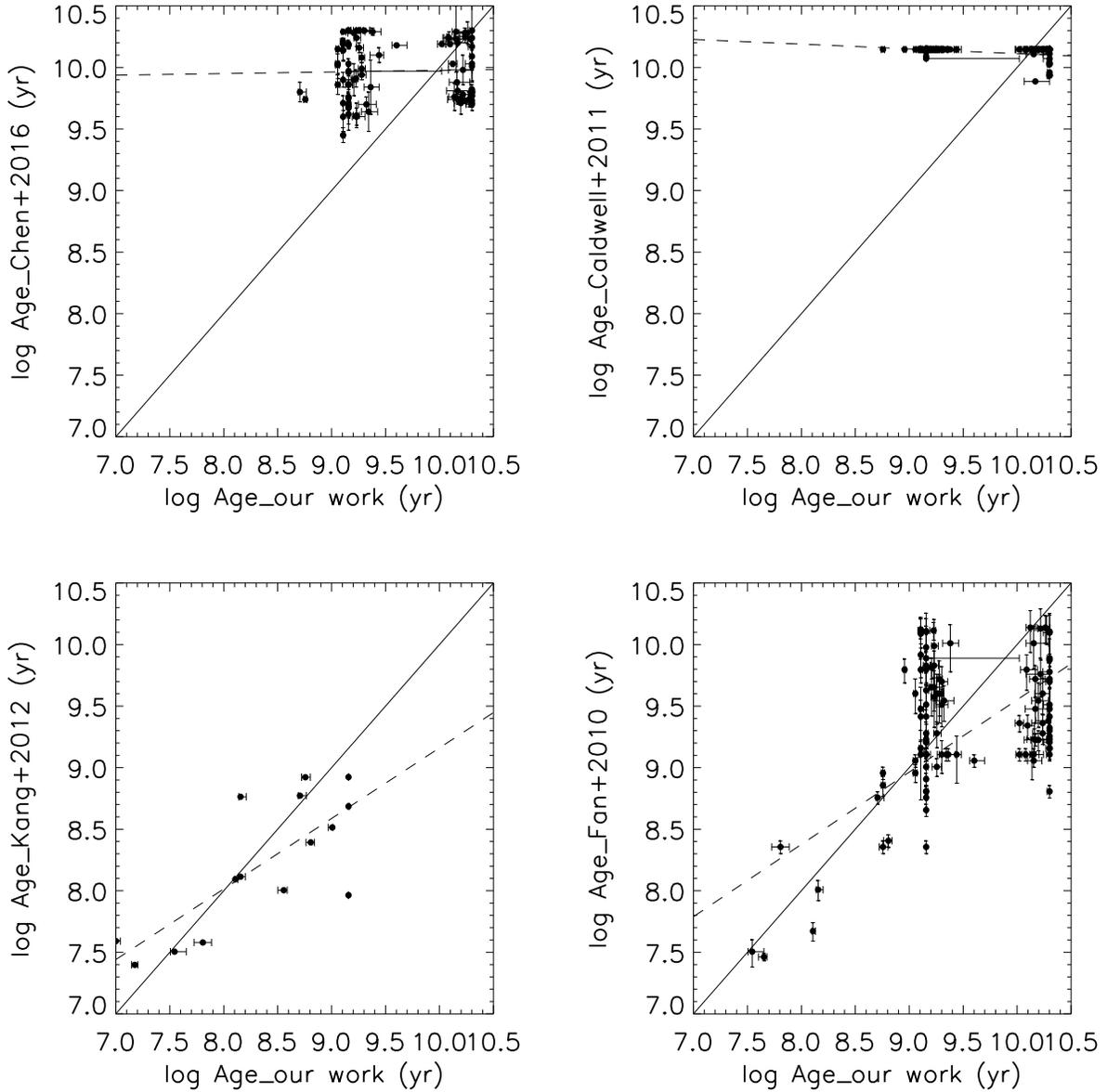}}}
  \caption{The same as Figure~\ref{fig1} but for comparisons of the ages 
    fit with Padova 2000 evolutionary track and \citet{chab03} IMF of 
    \citet{bc03} model and that from literature works: 
    \citet{cbq} (\sc Top Left Panel); \citet{cw11} (\sc Top Right Panel); 
    \citet{wang10} (\sc Bottom Left Panel); \citet{fan10} (\sc Bottom Right 
    Panel). The dashed lines represent the best linear fits.}
  \label{fig3}
\end{figure}

\begin{figure}
  \resizebox{\hsize}{!}{\rotatebox{0}{\includegraphics{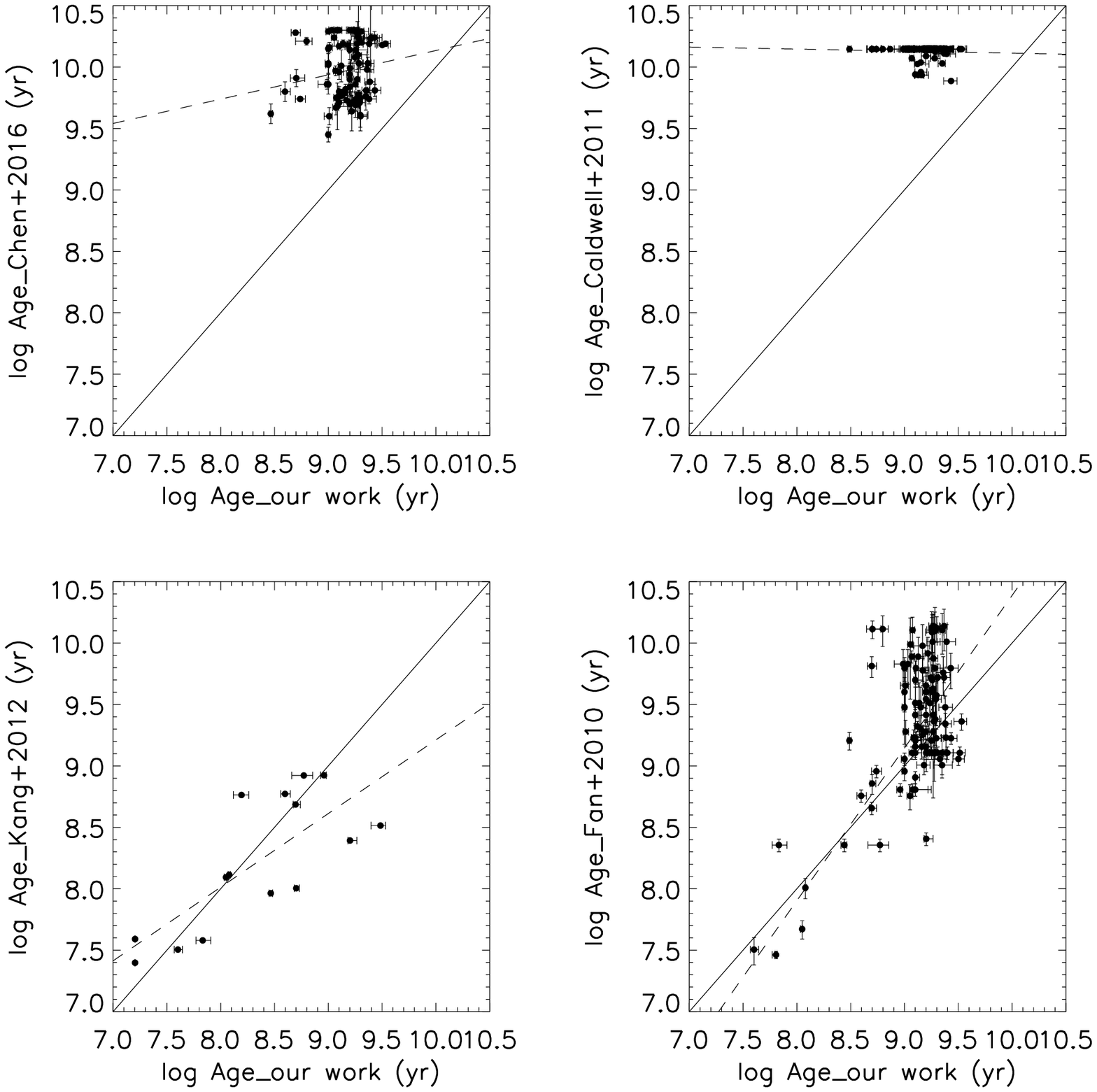}}}
  \caption{The same as Figure~\ref{fig2} but for {\sc galev} models of
    Kroupa IMF. The dashed lines represent the best linear fits.}
  \label{fig4}
\end{figure}

\clearpage
\begin{figure}
  \resizebox{\hsize}{!}{\rotatebox{0}{\includegraphics{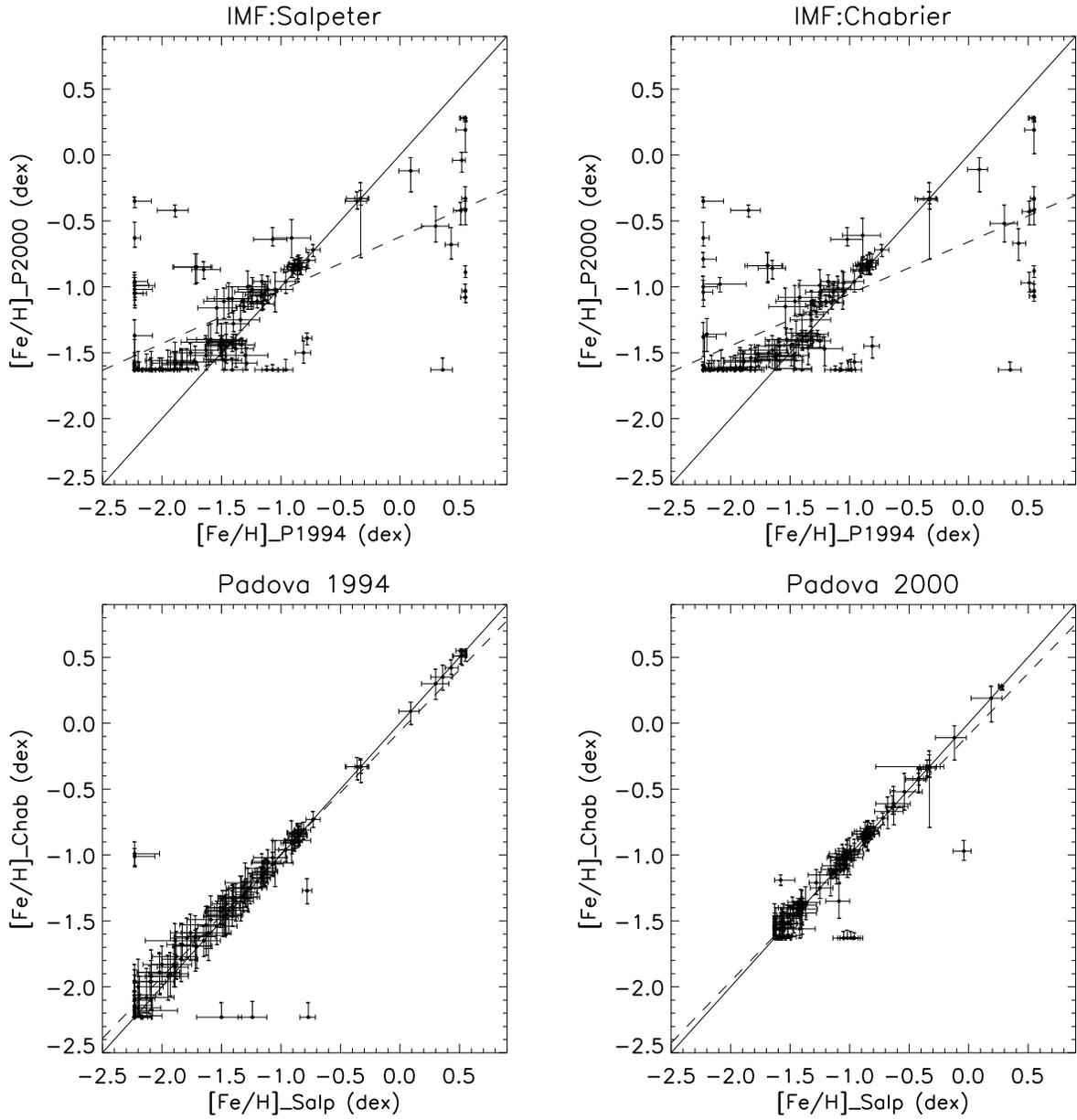}}}
  \caption{Comparisons of metallicity derived from the \citet{bc03} models
    of Padova 1994/Padova 2000 evolutionary tracks (\sc{Top Panels}) and 
    IMFs of \citet{chab03}/\citet{sal} (\sc{Bottom panels}). The photometry of
    all bands are used for the fitting. The dashed lines represent the best linear fits.}
  \label{fig5}
\end{figure}

\begin{figure}
  \resizebox{\hsize}{!}{\rotatebox{0}{\includegraphics{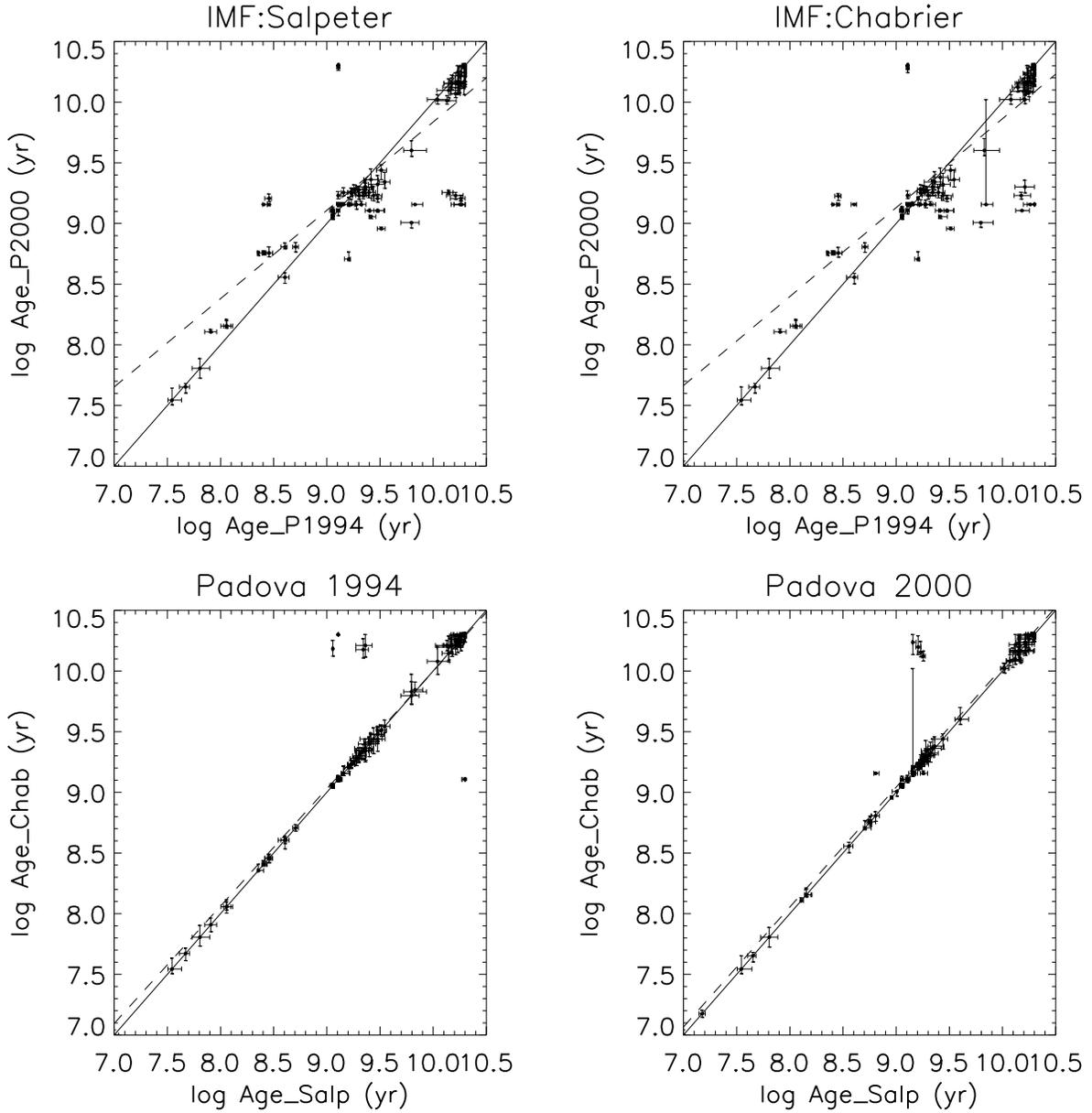}}}
  \caption{Same as Figure~\ref{fig5} but for the ages, which are derived from 
    evolutionary tracks of Padova 1994/Padova 2000 and IMFs of \citet{chab03}/
    \citet{sal} of the \citet{bc03} models. The photometry of all bands
    are used for the fitting. The dashed lines represent the best linear fits.}
  \label{fig6}
\end{figure}

\begin{figure}
  \resizebox{\hsize}{!}{\rotatebox{0}{\includegraphics{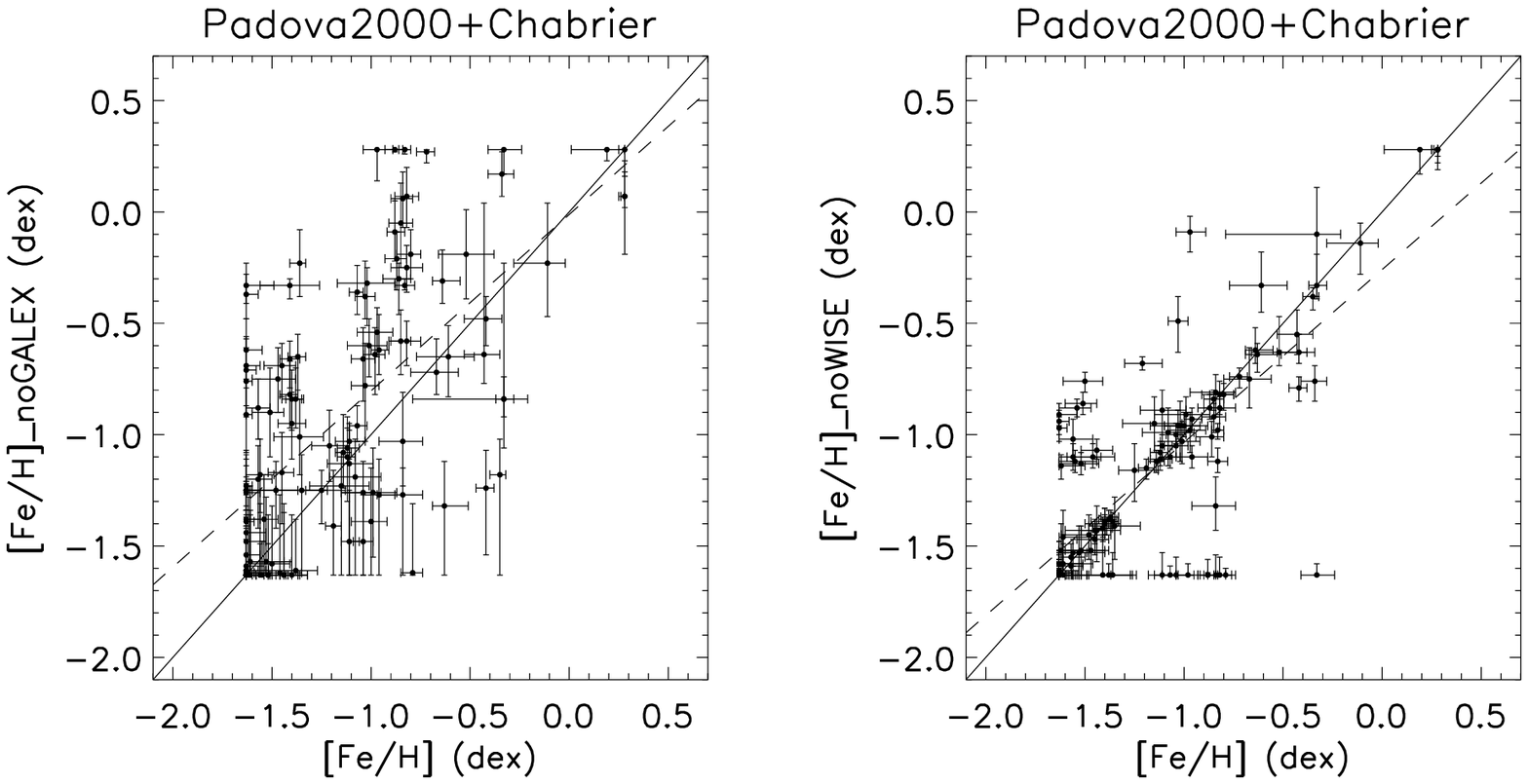}}}
  \caption{Comparisons of metallicity fitted from the BC03 models
    with photometry of all bands and that without {\it GALEX} data
    ({\sc Left Panel}) and that without WISE data ({\sc Right Panel}). 
    The Padova 2000 evolutionary track and \citet{chab03} IMF are applied. 
    The dashed lines represent the best linear fits.}
  \label{fig7}
\end{figure}

\begin{figure}
\resizebox{\hsize}{!}{\rotatebox{0}{\includegraphics{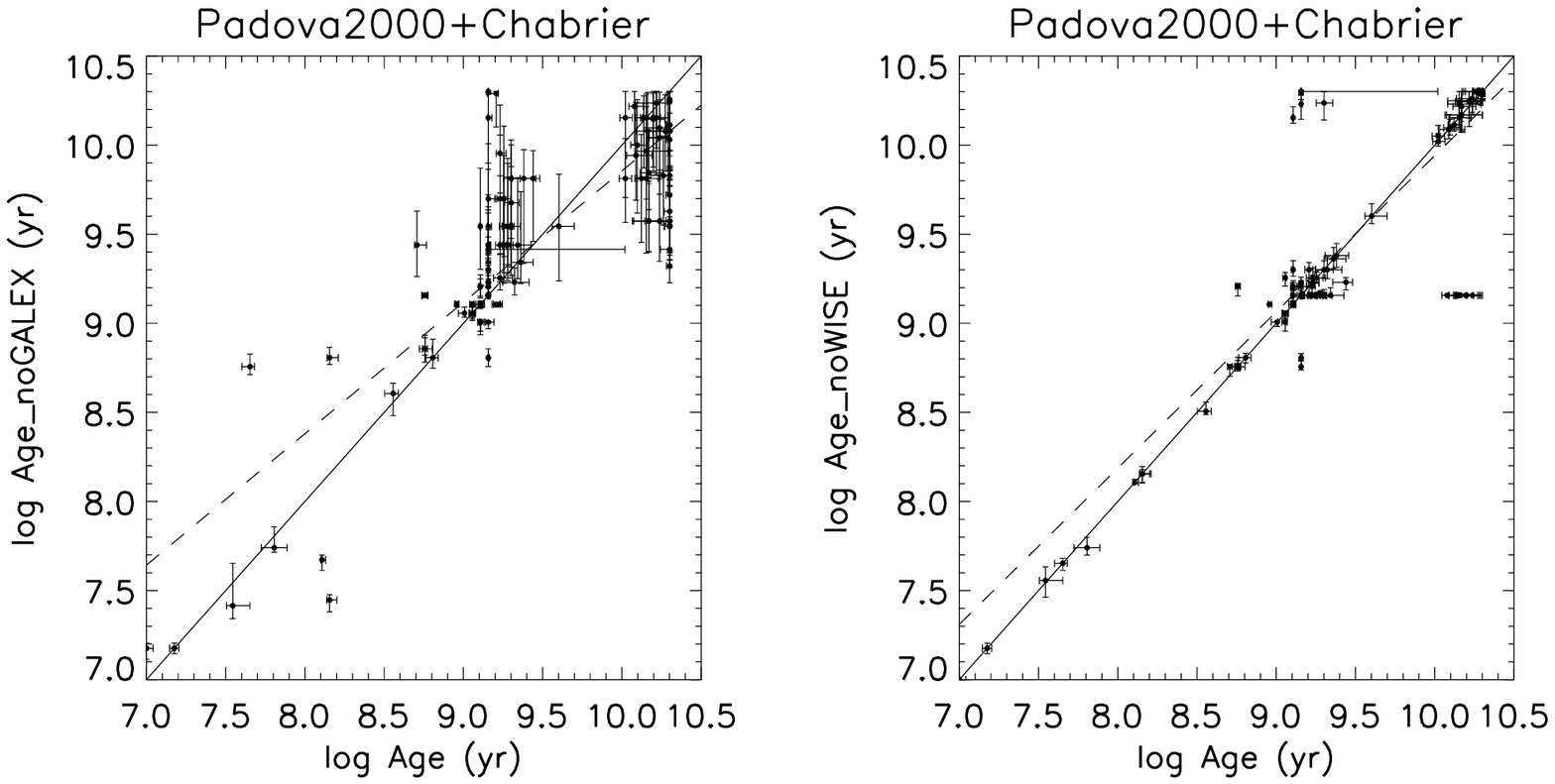}}}
\caption{Comparisons of ages fitted from the BC03 models 
  with photometry of all bands and that without {\it GALEX} data
  ({\sc Left Panel}) and that without WISE data ({\sc Right Panel}).
  The Padova 2000 evolutionary track and \citet{chab03} IMF are applied.
  The dashed lines represent the best linear fits.}
\label{fig8}
\end{figure}

\clearpage
\begin{figure}
  \resizebox{\hsize}{!}{\rotatebox{0}{\includegraphics{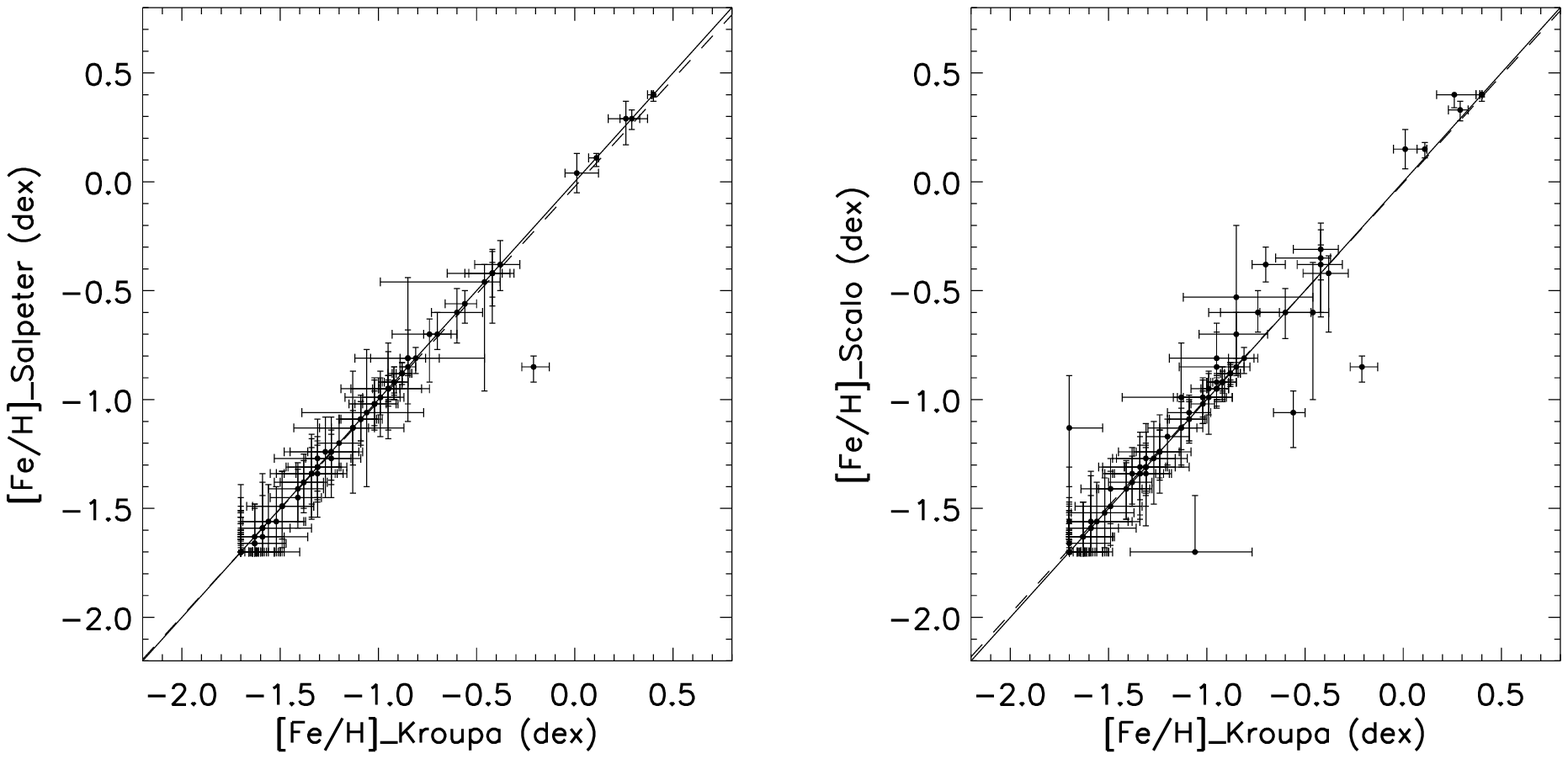}}}
  \caption{Same as Figure~\ref{fig5} but for {\sc galev} models of Kroupa/Scalo/\citet{sal} IMFs. The dashed lines represent the best linear fits.}
  \label{fig9}
\end{figure}

\begin{figure}
  \resizebox{\hsize}{!}{\rotatebox{0}{\includegraphics{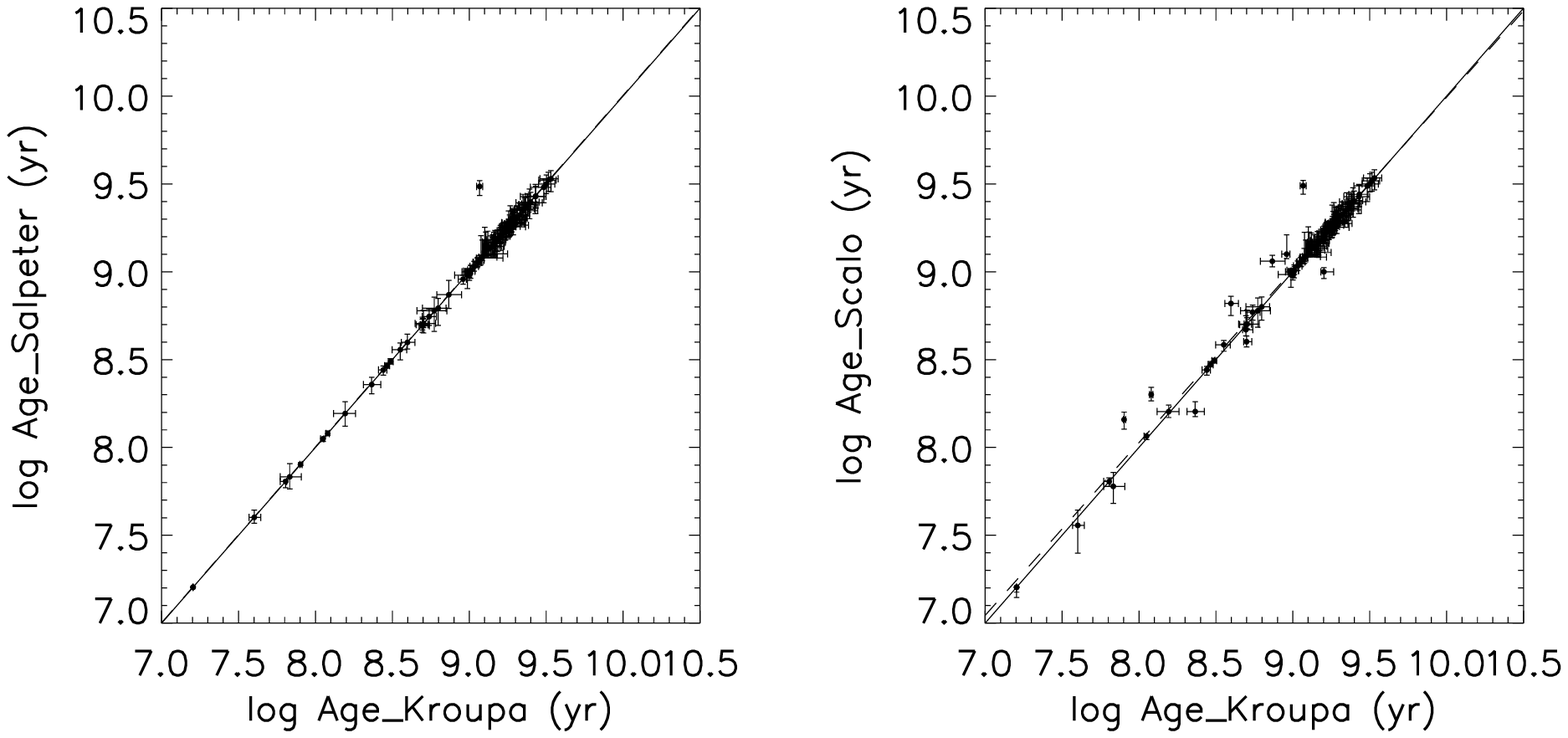}}}
  \caption{Same as Figure~\ref{fig6} but for {\sc galev} models with IMFs of Kroupa/Scalo/\citet{sal}. The dashed lines represent the best linear fits.} 
  \label{fig10}
\end{figure}

\begin{figure}
  \resizebox{\hsize}{!}{\rotatebox{0}{\includegraphics{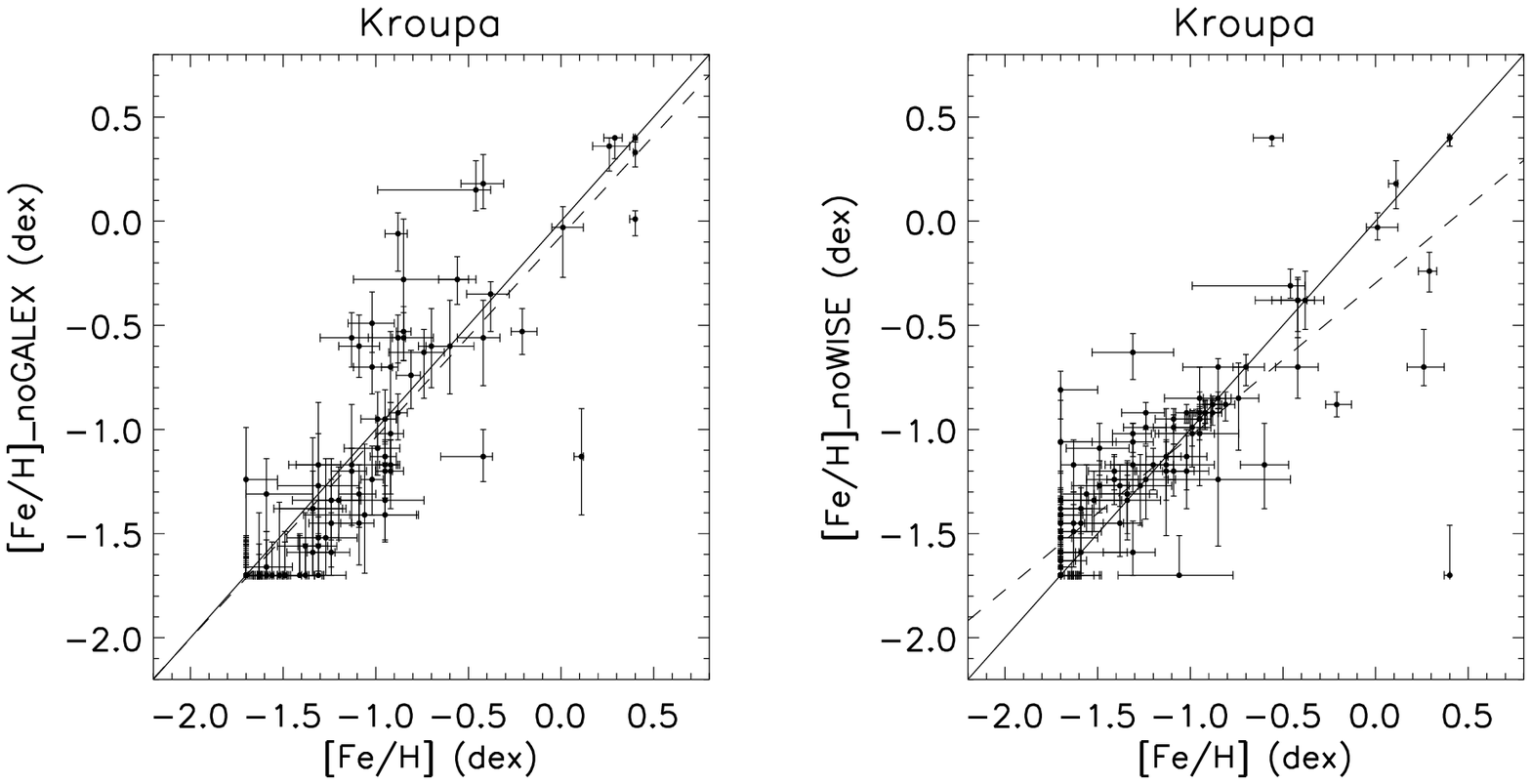}}}
  \caption{Same as Figure~\ref{fig7} bur for {\sc galev} models of Kroupa 
    IMF with photometry of all bands and that without {\it GALEX} data
    ({\sc Left Panel}) and that without WISE data ({\sc Right Panel}). 
    The dashed lines represent the best linear fits.}
  \label{fig11}
\end{figure}

\begin{figure}
  \resizebox{\hsize}{!}{\rotatebox{0}{\includegraphics{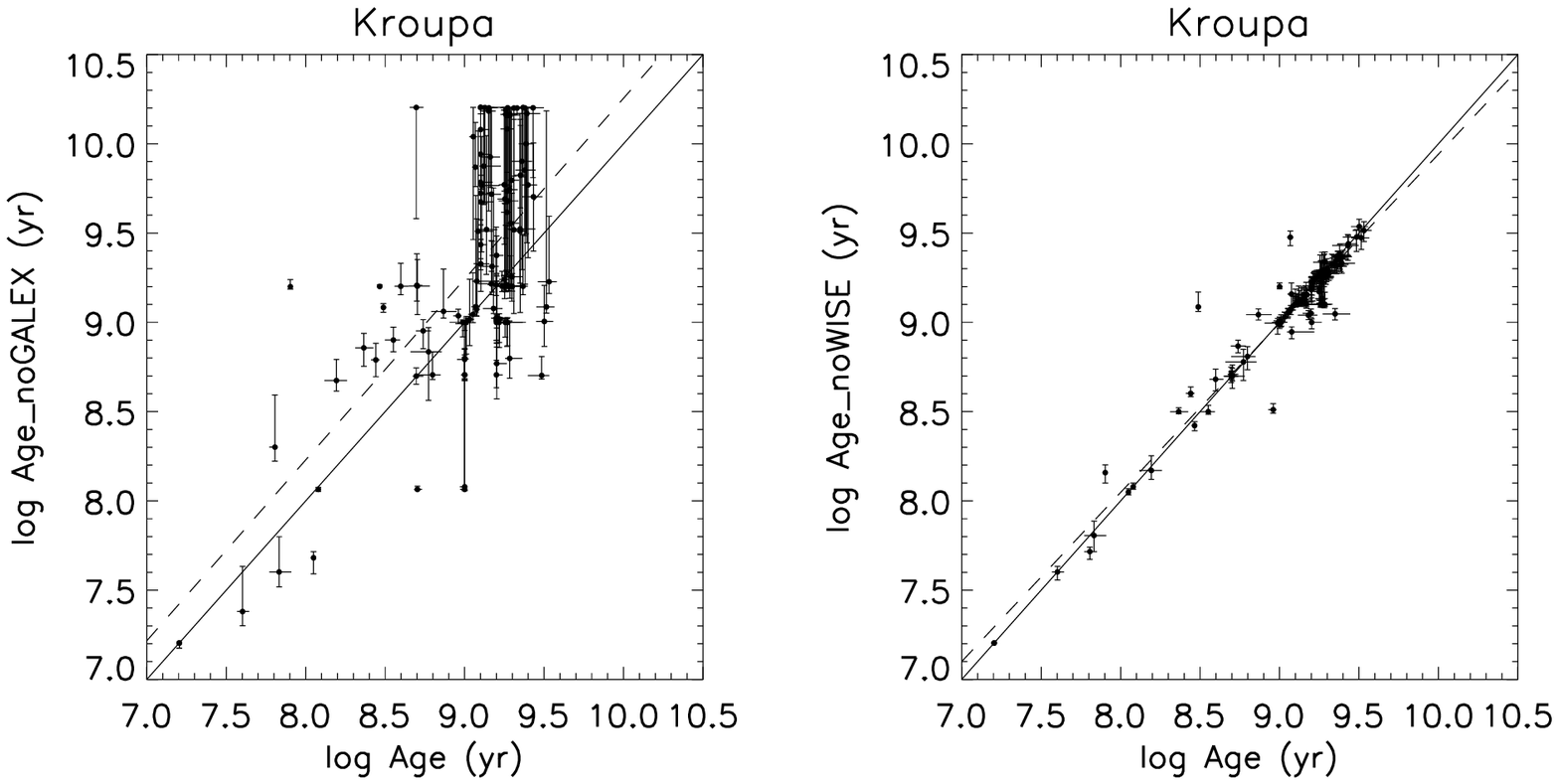}}}
  \caption{Same as Figure~\ref{fig8} but for {\sc galev} models of Kroupa IMF 
    with photometry of all bands and that without {\it GALEX} data 
    ({\sc Left Panel}) and that without WISE data ({\sc Right Panel}).
    The dashed lines represent the best linear fits.}
  \label{fig12}
\end{figure}

\label{lastpage}
\end{document}